\documentclass[aps,pra,twocolumn,superscriptaddress]{revtex4-1}
\usepackage{graphicx}
\usepackage{natbib}
\usepackage[breaklinks]{hyperref}

\newcommand{\hflev}[4]{\textit{#1}$_{#2/#3}$, \textit{F}=#4}
\usepackage{amsmath}

\begin{document}
\title{Photon bandwidth dependence of light-matter interaction}

\author{Matthias Steiner}
\affiliation{Center for Quantum Technologies, 3 Science Drive 2, Singapore 117543}
\affiliation{Department of Physics, National University of Singapore, 2
  Science Drive 3, Singapore 117542}
\author{Victor Leong}
\affiliation{Center for Quantum Technologies, 3 Science Drive 2, Singapore 117543}
\affiliation{Department of Physics, National University of Singapore, 2
  Science Drive 3, Singapore 117542}
\affiliation{Present address: Data Storage Institute, Agency for Science, Technology and Research (A*STAR), Singapore 138634, Singapore}
\author{Mathias Alexander Seidler}
\affiliation{Center for Quantum Technologies, 3 Science Drive 2, Singapore
  117543}
\author{Alessandro Cer\`{e}}
\affiliation{Center for Quantum Technologies, 3 Science Drive 2, Singapore
  117543}\
\author{Christian Kurtsiefer}
\email{christian.kurtsiefer@gmail.com}
\affiliation{Center for Quantum Technologies, 3 Science Drive 2, Singapore 117543}
\affiliation{Department of Physics, National University of Singapore, 2
  Science Drive 3, Singapore 117542}

\begin{abstract}
  We investigate the scattering of single photons by single atoms and, in particular, the dependence of the atomic dynamics and the scattering probability on the photon bandwidth. 
  We tightly focus the incident photons onto a single trapped~$^{87}${Rb} atom and use the time-resolved transmission to characterize the interaction strength.
  Decreasing the bandwidth of the single photons from 6 to~2 times the atomic linewidth, we observe an increase in atomic peak excitation and photon scattering probability.
\end{abstract}
\maketitle


\section{Introduction}
Hybrid quantum systems aim to overcome difficulties in implementing more
complex quantum information processing tasks with individual quantum
systems~\cite{Xiang2013,Kurizki2015} - 
a quantum network with solid-state systems as fast
processors, atomic systems as long-lived memories, and optical photons
interfacing spatially separated network nodes is an example~\cite{Kimble2008,Waks2009}.
There, control over the spectral and temporal properties of the exchanged photons is desirable for efficient information transfer
~\cite{Wilk:2007hm,Ritter2012}, but can be hard or impractical to implement.  
A common problem in hybrid quantum networks is a mismatch between the
characteristic time scale (or bandwidth) of photons emitted by one node with the optical transition of the receiving node~\cite{Akopian:2011,Ulrich2014,Siyushev2014,Jahn:2015,Meyer:2015}.
Understanding the role of the photon bandwidth in light-matter interaction is therefore important for the further development of hybrid networks. 

One realization of an atomic node of a quantum network is a single atom in free space coupled to a strongly focused mode~\cite{Sondermann2013}. 
Aside from the potential use in quantum networks, single atoms are also an ideal test bed to study fundamental properties of light-matter interaction. 
Following up on recent work on time-resolved scattering of photons with
exponentially rising and decaying profiles~\cite{Leong2016}, we report in this
work on the dependency of single photon scattering on their bandwidth.

\section{Theory}\label{sec:theory}
The interaction between single atoms and light has been studied extensively in semi-classical and fully quantized frameworks~\cite{Bufmmodeheckzlseziek1999,Enk2000,Enk2001,Enk2004,Domokos2002,Imamoglu:2008,Agio:2008,Wrigge2008,Heugel:2009,Leuchs:2009,Tey2009,Wang:2011,Leuchs:2012ps,Trautmann2016}.
Interesting recent results were the complete reflection of
light by a single atom~\cite{Agio:2008,Tey2009} and the complete
absorption of a single photon by reversing the spontaneous emission process from an excited two-level atom into
spherical harmonic modes described many decades ago~\cite{Weisskopf:1930jm,Imamoglu:2008,Leuchs:2009,Wang:2011}.

Practically important cases for traveling photons with a finite
duration have an exponential temporal envelope and consequently a Lorentzian
power spectrum 
\begin{equation}\label{eq:lorentzian}
\mathcal{L}\left(\omega\right)\propto
\frac{1}{\left(\omega-\omega_0\right)^2+\Gamma_\textrm{p}^2/4}
\end{equation}
of width $\Gamma_\textrm{p}$ centered at the atomic resonance frequency
$\omega_0$.
For a single photon Fock state with this power spectrum~(Eq.\ref{eq:lorentzian}),
 the scattering probability~$\epsilon$ is obtained by solving Eq.~18 and Eq.~21 in Ref.~\cite{Wang:2011},
\begin{equation}\label{eq:tx}
  \epsilon =
  4\Lambda \left(1-\Lambda \right) \frac{\Gamma_0}{\Gamma_0+\Gamma_\textrm{p}}\,,
\end{equation}
where $\Gamma_0$ is the width of the atomic line of interest and $\Lambda \in [0,1]$ is the spatial overlap between the excitation and the atomic dipole mode~\cite{Sondermann2013,Leong2016}. 
Photons of different temporal envelopes can have identical power
spectra -- for example, both exponentially rising and falling temporal envelopes lead
to a Lorentzian power spectrum. In the work presented here, we consider an
exponential decaying envelope $P_\textrm{p}(t)$ for the photon in the time domain, 
\begin{equation}\label{eq:photons_decay}
  P_\textrm{p}(t) = \Gamma_\textrm{p} \Theta(t) \exp{\left(-\Gamma_\textrm{p} t\right)}\,,
\end{equation}
where $\Theta(t)$ is the Heaviside step function.
For such a photon the probability~$P_\textrm{e}(t)$ of finding a two-level atom in the excited state is given by~\cite{Leong2016,Leuchs:2009}
\begin{equation}\label{eq:P_e_down}
  P_{\textrm{e}}(t)=
        \frac{4\Lambda \Gamma_0 \Gamma_\textrm{p} }{\left(\Gamma_0-\Gamma_\textrm{p}\right)^2}  \Theta(t)\left[ \exp{\left(-\frac{1 }{2}\Gamma_0 t\right)}- \exp{\left(-\frac{1}{2}\Gamma_\textrm{p} t\right)} \right]^2\,,
\end{equation}
with an atomic peak excitation probability
\begin{equation}\label{eq:pemax}
 P_\textrm{e,max} = 4 \Lambda \left( \frac{\Gamma_\textrm{p}}{\Gamma_0}\right)^{\frac{\Gamma_0+\Gamma_\textrm{p}}{\Gamma_0-\Gamma_\textrm{p}}}\,.
\end{equation}
According to this model, the highest atomic
peak excitation probability is reached if the bandwidth of the incident photon matches the atomic linewidth, $\Gamma_\textrm{p}=\Gamma_0$,
while the highest scattering probability
is obtained for very narrowband excitation, $\Gamma_\textrm{p} \rightarrow 0$~(Eq.~\ref{eq:tx}) .

\section{Experiment}
To test the model presented in Section~\ref{sec:theory}, we prepare single photon states by heralding on
a time-correlated photon-pair from a parametric process in a cold atomic
ensemble~\cite{Chaneliere:2006,Srivathsan:2013}. 
These photons are 
focused onto a single atom trapped in a far-off-resonant optical dipole trap.
From the fraction of photons that are scattered out of this focused
excitation mode we can accurately obtain the 
the transient atomic excitation~\cite{Piro:2010js,Sandoghdar:2012,Brito:2016,Leong2016}. 

\begin{figure}
\centerline{\includegraphics[width=\linewidth]{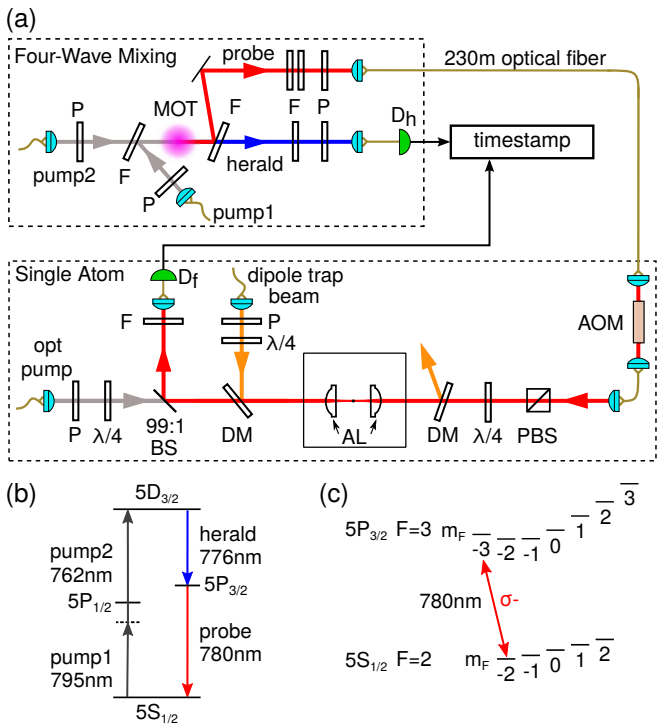}}
  \caption{\label{fig:setup}
    \textbf{(a)}~Optical setup to prepare heralded single photons at 780\,nm which are tightly focused on a single trapped atom. 
    Four-wave mixing: two co-aligned pump fields (pump1 at 795\,nm and pump2 at 762\,nm) generate herald (776\,nm) and probe (780\,nm) photon pairs in a cold cloud of $^{87}$Rb atoms.
    The detection of a herald photon at D$_\textrm{h}$ signals the presence of a single photon in the probe mode. 
    Single Atom: a $^{87}${Rb} atom is trapped at the focus of a confocal aspheric lens pair~(AL) with an optical dipole trap.
    D$_\textrm{h}$, D$_\textrm{f}$: avalanche photodetectors, P: polarizer, F: interference filter, $\lambda$/2, $\lambda$/4: half- and quarter-wave plate, (P)BS: (polarizing) beam splitter, DM: dichroic mirror, AOM: acousto-optic modulator, MOT: magneto-optical trap.
    \textbf{(b)}~Level scheme of the FWM process.
    \textbf{(c)}~Level scheme of the single  $^{87}$Rb atom in the dipole trap.}
\end{figure}

The experimental setup is shown in Fig.~\ref{fig:setup}(a).
A single atom is loaded from a cold ensemble 
in a magneto-optical
trap~(MOT) into a far-off-resonant optical dipole trap~\cite{Tey:2008} formed
by the tight focus of a light beam prepared by an aspheric lens 
(numerical aperture 0.55). The dipole laser (980\,nm, $42$\,mW, circular
polarization) provides a trap depth of approximately $2$\,mK.
Once trapped, the atom undergoes molasses cooling and is optically pumped into the 5\hflev{S}{1}{2}{2}, $m_F$=-2 state.
A bias magnetic field of $0.7$\,mT is applied along the optical axis.

We obtain heralded single photons from
correlated photon pairs generated by four-wave-mixing (FWM) in a cloud of cold
$^{87}$Rb atoms. This atomic cloud is
repeatedly cooled and refilled by a MOT for 140\,$\mu$s, followed by a photon
pair generation interval of~10\,$\mu$s.
During the photon pair generation, two pump beams with wavelengths 795\,nm and
762\,nm drive a transition from 5\hflev{S}{1}{2}{2} to
5\hflev{D}{3}{2}{3}~[Fig.~\ref{fig:setup}(b)]. A parametric
conversion process along a cascade decay channel generates time-ordered photon pairs. 
We collect herald~(776\,nm) and probe~(780\,nm) photons via interference filters into single mode fibers~\cite{Franson:1992cl,Srivathsan:2014jx,Gulati:2014}.
Detecting photons at 776\,nm then heralds single photons at 780\,nm, resonant with the $^{87}$Rb D2 transition.
Due to collective effects in the atomic ensemble, the bandwidth of the probe
photons is broader than the natural linewidth ($\Gamma_0/2\pi=6.07$\,MHz)
of the $5P_{3/2}-5S_{1/2}$ transition~\cite{Jen:2012,Srivathsan:2013}. 
We employ these effects to tune the bandwidth $\Gamma_\textrm{p}$ of the probe photons
over a range of 6 to 2~$\Gamma_0$ by controlling
the number and density of atoms in the cold ensemble via the gradient of the magnetic quadrupole field during the MOT phase. 

The probe photons are sent to the single atom setup via a 230\,m long optical
fiber. To suppress unheralded photons from the FWM setup, an acousto-optic
modulator~(AOM) acts as a switch 
between the photon
source and the single atom that is opened by a heralding event on
detector~D$_\textrm{h}$ for 600\,ns.  Optical and electrical delays are set such that the probe photon passes the AOM in the center of this interval.
This AOM also compensates for the 72\,MHz shift of the atomic resonance frequency caused by the bias magnetic field and the dipole trap.
Probe photons are focused onto the atom by the first aspheric lens. The
transmitted probe mode is then collimated again by a second aspheric lens,
subsequently coupled into a single-mode fiber, and sent to the forward
detector~D$_\textrm{f}$. In previous experiments we determined the spatial overlap of the probe mode with the atomic dipole mode to be $\Lambda\approx0.033$~\cite{Leong2016}.
\begin{figure}
 \centerline{\includegraphics[width=\linewidth]{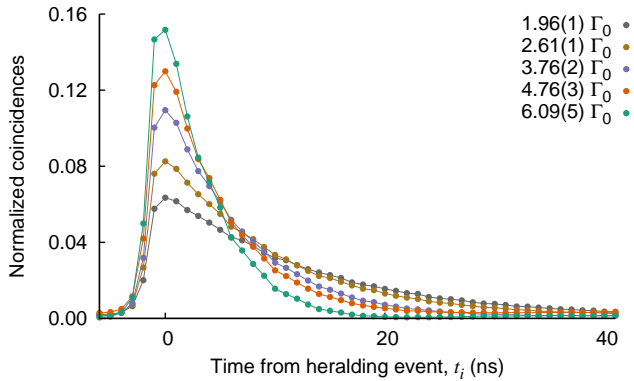}}
  \caption{\label{fig:photons}
    Coincidence histograms between heralding detector D$_\textrm{h}$ and forward detector D$_\textrm{f}$ for probe photons with five different bandwidths. 
    The bandwidths, shown in the figure legends, are determined by fitting the histograms to Eq.~\ref{eq:photons_decay}.
    Each histogram is normalized to the heralding~efficiency~$\eta_\textrm{f}$.
    Error bars are smaller than symbol size (one standard deviation
    of propagated Poissonian counting uncertainties). 
    Detection times are offset 
    to account for delays introduced by electrical and optical lines.
    }
\end{figure}

For each 
bandwidth setting, we obtain a coincidence histogram by sorting the detection
times~$t_\textrm{i}$ at~D$_\textrm{f}$ with respect to the 
heralding event into $\Delta t=1$\,ns wide time bins. 
Figure~\ref{fig:photons} shows reference
histograms~$G_\textrm{0}\left(t_\textrm{i}\right)$ with no trapped atoms, normalized to the heralding efficiency~$\eta_\textrm{f}= \sum_i  G_\textrm{0}\left(t_\textrm{i}\right)$ over the time interval $-10\,\textrm{ns}\leq t_{i} \leq 100\,\textrm{ns}$.
The time-ordering of herald and probe photons leads to an asymmetric
exponentially decaying profile, from
which we extract the corresponding photon bandwidth~$\Gamma_\textrm{p}$  by fitting to Eq.~\ref{eq:photons_decay} within the range $2\,\textrm{ns}\leq t_{i} \leq 100$\,ns. 
\begin{figure}
  \centering
  \includegraphics[width=\linewidth]{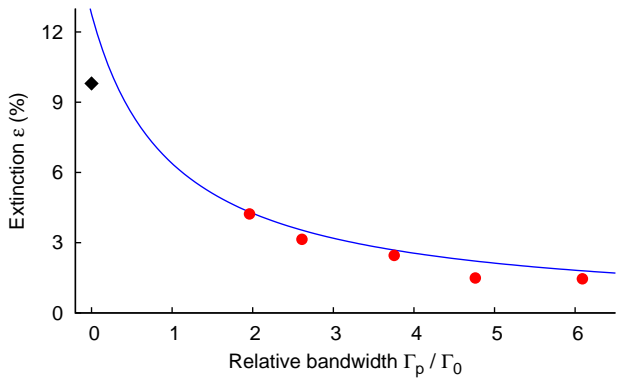}
  \caption{\label{fig:tx}
    Extinction~$\epsilon$ of probe photons with different bandwidth~$\Gamma_\textrm{p}$ (circles). For comparison we include the observed extinction of weak coherent field~(diamond). 
    Solid line:  Eq.~\ref{eq:tx} with $\Lambda=0.033$. Error bars are smaller than symbol size (one standard deviation
    of propagated Poissonian counting uncertainties).
  }
\end{figure}

When the atom is trapped, we record a corresponding set of probe
histograms~$G(t_\textrm{i})$ in a similar way.
From this, we can determine the extinction via
\begin{equation}
\epsilon=  1 -\sum\limits_i G\left(t_\textrm{i}\right) / \sum\limits_i
G_\textrm{0}\left(t_\textrm{i}\right)
\end{equation}
to characterize the photon-atom interaction. Both summations are performed over the time interval $-10\,\textrm{ns}\leq t_{i} \leq 100\,\textrm{ns}$.

\section{Results}
The results of the extinction measurements for bandwidth values between $6$
to $2\Gamma_0$ are shown in Fig.~\ref{fig:tx}. Consistent with the model
(Eq.~\ref{eq:tx}), 
the extinction increases for narrower photon bandwidths.
Since our heralded photon source can not efficiently prepare photons with a
bandwidth below $2\Gamma_0$,  
we use 100\,ms long pulses of laser light in a coherent state to simulate photons with $\Gamma_\textrm{p} \rightarrow 0$.
The observed extinction of the weak coherent field~(Fig.~\ref{fig:tx}, diamond) is larger than the extinction of the heralded photons, but deviates considerably from the model.
We attribute 
this mostly to linewidth broadening of the atomic transition:  
when tuning the frequency of the coherent probe field across the atomic resonance, we observe a linewidth of $2\pi\cdot10.6(6)\,\text{MHz}=1.7(1)\Gamma_0$ in the transmission spectrum.
This broadening can be  
caused by intensity fluctuations of the dipole trap field, the atomic motion of the atom, and the probe laser linewidth~\cite{Chin2016}.
The heralded photons from FWM have a larger bandwidth, so their scattering probability is less susceptible to this broadening of the atomic transition. 
\begin{figure}
  \centering
  \includegraphics[width=\linewidth]{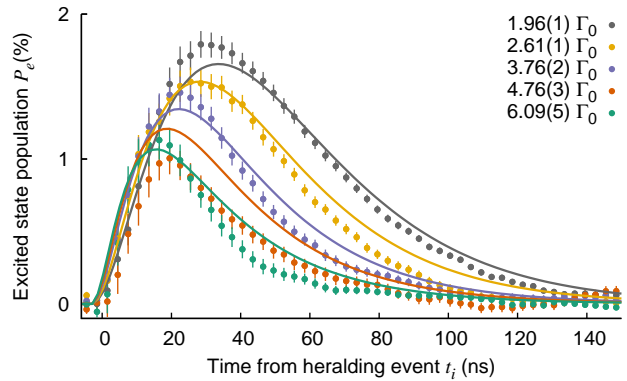}
  \caption{\label{fig:pe}
    Temporal evolution of the atomic excited population~$P_\textrm{e}(t)$ obtained from the time-resolved changes in transmission detection rate (see Eq.~\ref{eq:P_e_dot}).
    Narrowband photons lead to stronger and longer lasting atomic excitation, in agreement with Eq.~\ref{eq:P_e_down}~(solid lines).
    Error bars represent one standard deviation of the distributions obtained by a Monte-Carlo method which assumes Poissonian statistics for the detection rates.}
\end{figure}

\begin{figure}
  \centering
  \includegraphics[width=\linewidth]{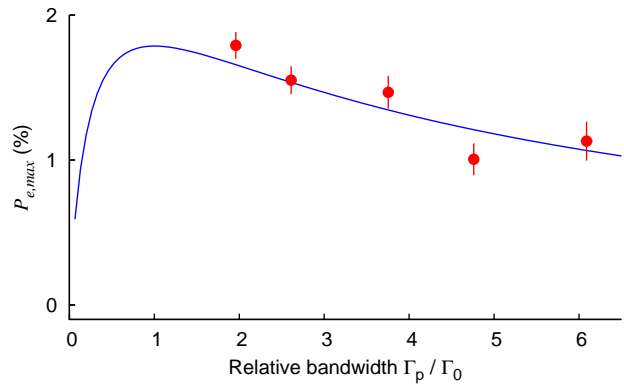}
  \caption{\label{fig:pe_max}
  Atomic peak excitation~$P_\textrm{e, \textrm{max}}(t)$, extracted from the experimental~$P_\textrm{e}(t)$ curves shown in Fig.~\ref{fig:pe}.
  Solid line: Eq.~\ref{eq:pemax}.}
\end{figure}
Aside from the scattering probability, we also determine the temporal
evolution of the atomic excited population $P_\textrm{e}(t)$ during the
scattering process.
Any change of the forward detection rate~$
\delta\left(t_\textrm{i}\right)= \left(G_\textrm{0}(t_\textrm{i}) -
G(t_\textrm{i})\right)/\left(\eta_\textrm{f} \Delta t\right)$ is directly related to a change
$\dot{P}_\textrm{e}(t)$ of the atomic population~\cite{Leong2016} via the rate
equation
\begin{equation} \label{eq:P_e_dot}
  \dot{P}_\textrm{e}(t) = \delta(t) - (1-\Lambda)\Gamma_0 P_\textrm{e}(t)\,.
\end{equation}

Thus, $P_\textrm{e}(t)$ can be obtained by integrating Eq.~\ref{eq:P_e_dot}
with a relatively low experimental uncertainty. Figure~\ref{fig:pe} shows the
evolution of the resulting $P_\textrm{e}(t)$, 
agreeing very well with Eq.~\ref{eq:P_e_down} (solid lines) for narrowband photons,
but exhibit a stronger deviation from the model as the bandwidth is increased.
This deviation 
is likely due to the imperfect photon profiles on the sharp rising edge 
(Fig.~\ref{fig:photons}) compared to the ideal asymmetric exponentially decaying profile. 

The peak excitation probability~$P_\textrm{e,max}$ for each bandwidth in
Fig.~\ref{fig:pe} is shown in Figure~\ref{fig:pe_max}, and is in good agreement with Eq.~\ref{eq:pemax}. 
Comparing results for narrowband ($\Gamma_\textrm{p}= 1.96(1)\Gamma_0$) and broadband
($\Gamma_\textrm{p}=6.09(5)\Gamma_0$) photons, we observe an increase in the
peak excitation by a factor~$1.5(2)$.
This relative increase is smaller than the relative increase by a factor of
2.6(4) of the extinction between these two bandwidths, i.e.,  the atomic peak
excitation has a weaker dependence on the photon bandwidth than the scattering
probability over our accessible bandwidth range.  

\section{Discussion and summary}
While often a semiclassical description of light atom-interaction can
explain observations with sufficient accuracy, it is well known that the
photon statistics of the incident light can affect the atomic 
excitation dynamics~\cite{Georgiades:1995,Domokos2002,Wang:2011, Strauss:2016}. 
Full quantum models of the light-matter interaction predict different scattering probabilities for single photons compared to coherent fields with equal bandwidth and a mean photon number of one~\cite{Domokos2002,Wang:2011}.  
As (hybrid) quantum networks in an information processing scenario are likely
to operate by exchanging single photons, testing light-matter interfaces with
single photons -- as opposed to weak coherent fields -- becomes important. 
For the experimental parameters in the work presented here ($\Lambda=0.033$,
$\Gamma_\textrm{p}=2\Gamma_0$), the expected difference in scattering
probability is only 0.05\%, and thus within our experimental uncertainties. 
However, already a slight improvement of experimental parameters parameters
should make the difference in scattering resolvable, assuming similar
experimental uncertainties. 
For $\Lambda=0.1$ and $\Gamma_\textrm{p}=\Gamma_0$, quite within
experimentally reachable range~\cite{Alber2016}, we expect $\epsilon=18.0\%$ for single photon
excitation, but $\epsilon=17.1\%$ for a coherent field. 

In summary, we find that the role of the photon bandwidth in the scattering
process with a single atom is well described by a relatively simple  excitation
model with a fully quantized field description~\cite{Wang:2011}. Tuning the
photon bandwidth from 6 to 2~$\Gamma_0$, we observe an increase in the
scattering probability as well as in the atomic peak excitation. 
Notably, the relative increase in the scattering probability is larger than in the atomic peak excitation.

\section*{Funding}
Ministry of Education in Singapore (AcRF Tier 1);  
National Research Foundation, Prime Minister's office (partly under grant No. NRF-CRP12-2013-03); 
M. Steiner acknowledges support by the Lee Kuan Yew Postdoctoral Fellowship.

\bibliographystyle{apsrev4-1}
\bibliography{tx_safwm}

\end{document}